\documentstyle[amstex,aps]{revtex}
\begin{document}
\title{Smoothed Particle Hydrodynamics for Relativistic Heavy Ion Collisions }
\author{C.E. Aguiar$^{1}$, T. Kodama$^{1}$, T. Osada$^{2}$ and Y. Hama$^{2}$}
\address{$^{1}$Instituto de F\'{i}sica, Universidade Federal do Rio de Janeiro\\
$^{2}$Instituto de F\'{i}sica, Universidade de S\~{a}o Paulo, S\~{a}o Paulo}
\maketitle

\begin{abstract}
The method of smoothed particle hydrodynamics (SPH) is developped appropriately
for the study of relativistic heavy ion collision processes. In order to
describe the flow of a high energy but low baryon number density fluid, 
the entropy is taken as the SPH base. We formulate the method in 
terms of the variational principle. Several examples show that the method 
is very promising for the study of hadronic flow in RHIC physics.
\end{abstract}

\section{Introduction}

Hydrodynamic descriptions of high energy hadronic and nuclear collisions
have a rather long history\cite{Landau}. Although, from theoretical point of
view, it is not a trivial matter to justify their validity, they have been
successful in reproducing certain features of these processes, such as the
energy dependence of the average multiplicity and the transverse-energy
distributions. More recently, relativistic fluid dynamics have become an
important tool for the analysis of relativistic heavy-ion-collision
processes\thinspace(for example, \cite{Strottman,Csernai,Stöcker} and
references there in). In these processes, the nuclear matter is expected to
be compressed and heated up close to those states of the matter realized in
the Big Bang Era of the Universe. Some laboratory data of these fascinating
processes have already been obtained in the series of CERN experiments\cite
{CERN}, and data at even higher temperatures and densities will soon be
available in the forthcoming RHIC experiments.

The relativistic hydrodynamics is a description based on {\it local}
conservation laws, together with the hypothesis of local thermodynamical
equilibrium. The conservation laws are written in terms of the
four-divergence of the energy-momen\-tum tensor. The resulting system of
equations is highly nonlinear and analytical solutions are only available
for some very particular and limited configurations and equations of state 
\cite{Landau,Bjorken}. Thus numerical approaches are resorted to but they
usually depend on some sophisticated techniques specific to some symmetry
involved in the problem. When no symmetry is involved, these methods become
computationally very expensive. However this is exactly the case when we are
challenged to a realistic three dimensional simulation for nuclear-collision
processes. There, we expect no geometrical symmetry so that a full 3D
calculation is required.

One basic point in the hydrodynamic approach of relativistic nuclear
collisions is that its principal ingredients, i.e., the equation of state of
the matter and the initial conditions for the dynamics are not quite well
known. On the contrary, we apply the hydrodynamic models to infer these
informations on the properties of the matter in such a highly condensed and
excited state. Thus we need to perform many hydrodynamic-model calculations
for different equations of state and initial conditions to compare with the
experimental data. In such a process, we actually don't need the very
precise solution of hydrodynamic equations, but a general flow pattern which
characterizes the final configuration of the system as a response to a given
set of equation of state and initial conditions. We are not interested in,
and probably neither able to analyze at least in the present stage, any very
precise local feature (for example sound ripples, small local perturbations,
etc.) in these models. Extremely local properties, therefore, should be
averaged out without spoiling the general flow pattern. This is particularly
the case when one considers the questionable validity of the local thermal
equilibrium in the problem of interest here. In short, for the study of
hydrodynamic models of relativistic nuclear collisions, we prefer a rather
simple scheme of solving hydrodynamic equations, not unnecessarily too
precise but robust enough to deal with any kind of geometry. From this point
of view, it is stressed in Ref. \cite{Variation} the advantage of
variational approach to the relativistic hydrodynamics.

Among many numerical approaches, the Smoothed Particle Hydrodynamics (SPH) 
\cite{Lucy,Monaghan1} fits perfectly in with the variational formalism. As a
consequence, this approach presents many desirable profiles of the
variational approach, such as simplicity and robustness with respect to the
changes of geometry, as well as the possibility of smoothing out undesirable
local degrees of freedom. Furthermore, the SPH parametrizes the matter flow
in terms of discrete Lagrangian coordinates (called ``particles'') attached
to some conserved quantity, such as baryon number. In this aspect, the SPH
method is the best to the studies of relativistic nuclear collisions, where
a extremely compressed and high-temperature hadronic matter expands into a
very large space region. This is one of the principal advantages of the SPH
over the other space-fixed grid algorithms when applied to the study of RHIC
physics. The SPH algorithm was first introduced for astrophysical
applications and used by now in several calculations, such as fragmentation
of asteroids, supernova explosions, collision of neutron stars, etc. Several
studies of the SPH algorithm for the ultrarelativistic regime of
hydrodynamics have been done\cite{ChowMonag,Zurek,Siegler,Thacker}.

However, some specific aspects of the relativistic heavy ion collision
processes deserve attention when applying the SPH. One of them is that, in
the ultrarelativistic regime of central collisions, a large fraction of the
incident energy is converted into produced particles. In particular, in the
mid-rapidity region of central collisions, the most of these energies are in
the form of produced pions and only a very small portion is carried by
baryons. This is a very unfavorable situation to the conventional
formulation of the SPH algorithm, where particles are associated to the
matter defined by the conserved quantity, such as the baryon number. A
direct application of the SPH based on the conserved baryon number may fail
in the null baryon-number, pion-dominated region.

As mentioned above, the basic point of the SPH method is to introduce a set
of ``SPH - particles'' which follow the flow of the fluid. However, the
definition of a flow does not necessarily rely on the conserved quantity.
For example, according to Landau\cite{LL} a flow is defined in terms of the
local Lorentz frame in which the energy-momentum tensor becomes diagonal. In
this paper, we explore this aspect and formulate the SPH in terms of any
extensive quantities defined in the Landau comoving local frame. We derive
the relativistic SPH equations using the variational principle\cite
{Variation}, taking the matter flow as the variable. We argue that the
quantity we attribute to the SPH particles convenient for relativistic
heavy-ion collisions are the entropy of the fluid. In this way, we can
follow directly the entropy content and its change due to the dissipation
mechanism, for example, the shock wave. This is particularly interesting for
the system where the first order phase transition is present \cite
{Blaizot,Baym,Ruusk}.

Another specific aspect of relativistic heavy-ion collision processes is how
to set the initial conditions for the hydrodynamic motion. As mentioned
before, this is not a solved problem. However, generally speaking, the
hydrodynamic regime will be established from its initial excited state of
the microscopic QCD degrees of freedom, only after a certain relaxation time 
$\tau_{relax}$. This means that the hypersurface of the onset of the
hydrodynamic regime is characterized by a nearly constant local proper time
rather than by a constant global coordinate time $t$. For ultrarelativistic
regime, the difference becomes crucial. In the limit of very large initial
energy, the Bjorken scaling solution is expected to be a good first order
approximation at least for the longitudinal motion. Therefore, we shall
write the hydrodynamic equations in terms of the Bjorken scaling coordinates 
$\tau=\sqrt{t^{2}-z^{2}}$ and $\eta=1/2\ln(t+z)/(t-z)\;$, replacing $t$ and $%
z$, as far as the longitudinal motion is concerned. These variables are also
suitable for the setup of the initial conditions. Here, we show that the
variational approach is also useful for the derivation of the SPH equations
for an arbitrary curvilinear coordinate system.

We organize the present paper as follows. In Sec.II, we introduce the
relativistic variational formulation with the SPH parametrization of one of
the thermodynamical extensive quantities. In particular, we argue that the
entropy is a convenient extensive quantity for the application of
relativistic heavy-ion collisions. In this case, the change of entropy due
to dissipative processes should be taken into account. This is particularly
important in the presence of a shock wave or the first order phase
transition. For this purpose, we deduce the entropy based SPH equations in
the presence of bulk viscosity. In Sec.III, we apply our formulation for
several relativistic systems such as one- and three-dimensional Landau
model, the one presenting longitudinal Bjorken scaling solution plus
transverse expansion, and one-dimensional relativistic fluid with shock
waves. These results are compared with known solutions. Sec. IV is dedicated
to the conclusions and perspectives of the present work.

\section{Variational derivation of the Relativistic SPH equation}

\subsection{SPH Representation of Extensive Variables}

The non-relativistic SPH can be formulated in terms of the variational
principle. We show here that the relativistic SPH can also be formulated in
this way\cite{Variation}. This guarantees that the SPH coordinates $\left\{ 
\vec{r}_{a}\left( t\right) \right\} $are the optimal dynamical parameters to
minimize the SPH-model action. The other advantage of this approach is that
it automatically leads to the symmetrized form of the SPH equation to
conserve linear and angular momenta of the system\cite{Monaghan}. This is
the natural consequence of the Lorentz scalar nature of the action.

In the relativistic hydrodynamics, we assume that at any space-time point $%
x=\left\{ \vec{r},t\right\} $ there exists a local reference frame where the
energy-momentum tensor becomes diagonal and takes the form\cite{LL} 
\begin{equation}
T^{\mu\nu}\left( \vec{r},t\right) =\left( 
\begin{array}{cccc}
\varepsilon & 0 & 0 & 0 \\ 
0 & P & 0 & 0 \\ 
0 & 0 & P & 0 \\ 
0 & 0 & 0 & P
\end{array}
\right) ,
\end{equation}
where $\varepsilon$ and $P$ are the (proper) energy density and the pressure
of the fluid. From the hypothesis of local equilibrium, we assume that the
thermodynamical relations are valid in each local frame.

Let $A$ be an arbitrary thermodynamical extensive quantity of the fluid such
as baryon number, entropy or specific volume. The amount of $A$ contained in
the infinitesimal volume element $dV$ is denoted by $dA$ so that the
corresponding (proper) density $a$ is 
\begin{equation}
a=\frac{dA}{dV}\;.  \label{dA/dV}
\end{equation}
This is related to the corresponding density $a^{\ast}$ measured in the
space-fixed (calculational) frame as 
\begin{equation}
a^{\ast}\left( \vec{r},t\right) =\gamma a,
\end{equation}
where $\gamma$ is the local Lorentz factor associated to the flow.

In the SPH representation, we parametrize this density by the following
ansatz, 
\begin{equation}
a^{\ast}\left( \vec{r},t\right) =\sum_{i}\nu_{i}W\left( \vec{r}-\vec{r}%
_{i}\left( t\right) ;h\right) ,  \label{omega_r}
\end{equation}
where $W\left( \vec{r}-\vec{r}\,^{\prime};h\right) $ is a positive definite
kernel function peaked at $\vec{r}=\vec{r}\,^{\prime}$ with the
normalization 
\begin{equation}
\int d^{3}\vec{r}W\left( \vec{r}-\vec{r}\,^{\prime};h\right) =1\,.
\end{equation}
The parameter $h$ represents the width of the kernel. In the limit, $%
h\rightarrow0$, we have 
\[
\lim_{h\rightarrow0}W\left( \vec{r}-\vec{r}\,^{\prime};h\right) =\delta
^{3}\left( \vec{r}-\vec{r}\,^{\prime}\right) . 
\]
As will be seen later, it is convenient to choose an even function for $W.$
It can be a Gaussian function in $x=\left| \vec{r}-\vec{r}\,^{\prime}\right|
/h$, but in practice we often use the B-spline functions\cite{Monaghan}. The
role of $W$ with a finite value of $h$ is to introduce a sort of short-wave
length cut filter in the Fourier representation of the density $a^{\ast}$.
The total amount of $A$ of the system is obtained by integrating Eq.(\ref
{omega_r}) over the whole space, 
\begin{equation}
A_{tot}=\int d^{3}\vec{r}\,a^{\ast}\left( \vec{r},t\right) =\sum_{i}\nu_{i}
\end{equation}
Physically speaking, it is clear from the above expression that we are
replacing the system of continuous fluid by a collection of ``SPH
particles'' each of which carries $\nu_{i}$ portion of the extensive
quantity $A$.

The velocity of these particles are identified as the velocity of the fluid
at their position $\vec{r}_{i}(t),$ 
\begin{equation}
\vec{v}_{i}=\frac{d\vec{r}_{i}}{dt},  \label{drdt}
\end{equation}
so that the Lorentz factor of the $i$-th particle is given by $\gamma _{i}=1/%
\sqrt{1-\vec{v}_{i}^{2}}.$ If $A$ is a conserved quantity, then $\left\{
\nu_{i}\right\} $ should be constant in time. Then 
\begin{align*}
\frac{\partial a^{\ast}}{\partial t} & =-\sum_{i}\nu_{i}\vec{v}%
_{i}\cdot\nabla W\left( \vec{r}-\vec{r}_{i}\left( t\right) ;h\right) \\
& =-\nabla\cdot\sum_{i}\nu_{i}\vec{v}_{i}W\left( \vec{r}-\vec{r}_{i}\left(
t\right) ;h\right) \\
& =-\nabla\cdot\vec{j}_{A}\;,
\end{align*}
where 
\[
\vec{j}_{A}=\sum_{i}\nu_{i}\vec{v}_{i}W\left( \vec{r}-\vec{r}_{i}\left(
t\right) ;h\right) 
\]
is the current density of $A$. Thus, the continuity equation is satisfied by
the ansatz, (\ref{omega_r}), together with Eq.(\ref{drdt}). If $A$ is not
conserved, then $\nu_{i}^{\prime}s$ are not constant in time. In this case,
the continuity equation has the contribution from the time derivative of $%
\nu_{i}^{\prime}s$ as 
\begin{equation}
\frac{\partial a^{\ast}}{\partial t}+\nabla\cdot\vec{j}_{A}=\sum_{i}\dot{\nu 
}_{i}W\left( \vec{r}-\vec{r}_{i}\left( t\right) ;h\right) .
\end{equation}

We consider the set of time dependent variables $\left\{ \vec{r}%
_{i},i=1,...,n\right\} $ as the variational dynamical variables and their
equations of motion are determined by minimizing the action for the
hydrodynamic system. Here, $\left\{ \nu_{i}\right\} $ are not dynamical
variables and are determined by the initital conditions together with the
constraints for the variational procedure (see the later discussion in {\bf %
2.3.}).

It may seem that the spatial extension of the particle $i$ is identified to
that of the kernel $W\left( \vec{r}-\vec{r}_{i};h\right) $. However, the
density of $A$ at the position of the particle $i$ is given by 
\begin{equation}
a_{i}^{\ast}=\sum_{j}\nu_{j}W\left( \vec{r}_{i}-\vec{r}_{j}\right) ,
\label{a*}
\end{equation}
from Eq.(\ref{omega_r}). Therefore, we may define the ``specific volume'' $%
V_{i}$ of the extensive quantity $A$ associated to the particle $i$ as 
\begin{equation}
V_{i}\equiv\frac{\nu_{i}}{a_{i}}=\frac{\gamma_{i}\nu_{i}}{\sum_{j}\nu
_{j}W\left( \vec{r}_{i}-\vec{r}_{j}\right) }\,.  \label{V}
\end{equation}

Any other extensive quantities carried by the particle $i$ can be calculated
easily. Let $o^{\ast}$ be the density of another extensive quantity, say $O, 
$ measured in the calculational frame. Then the amount of $O\,$carried by
the particle $i$ is 
\begin{equation}
\nu_{i}\left( O/A\right) _{i}=\nu_{i}\left( o/a\right) _{i}=\nu_{i}\left(
o/a\right) _{i}^{\ast}\,,  \label{O=Nu}
\end{equation}
so that the density distribution of $O$ in the calculational frame is
expressed as 
\begin{equation}
o^{\ast}\left( \vec{r},t\right) \rightarrow\sum_{i}\nu_{i}\left( o/a\right)
_{i}^{\ast}W\left( \vec{r}-\vec{r}_{i}\left( t\right) \right) .  \label{o*}
\end{equation}

\subsection{\protect\bigskip SPH Action}

The relativistic hydrodynamic equations can be obtained by the variational
principle for the action\cite{Variation} 
\begin{equation}
I=-\int d^{4}x\;\varepsilon\,,
\end{equation}
with respect to the matter density distribution, subjected to the constraint
expressed by the continuity equation. Here, for the sake of simplicity, we
discuss first the case of the Minkowsky metric. The argument can readily be
generalized for more general coordinate systems (see Sec.III). The
Lagrangian of the system is 
\begin{equation}
L=-\int d^{3}\vec{r}\;\varepsilon\,.
\end{equation}
We may consider $\varepsilon$ as the Lagrangian density in the space fixed
frame. Therefore, the corresponding Lagrangian for the system of SPH
particles can be taken to be 
\begin{align}
L_{SPH}\left( \left\{ \vec{r}_{i},\frac{d\vec{r}_{i}}{dt}\right\} \right) &
=-\sum_{i}\nu_{i}\left( \varepsilon/a^{\ast}\right) _{i}  \nonumber \\
& =-\sum_{i}\left( \frac{E}{\gamma}\right) _{i},  \label{L_SPH}
\end{align}
where $\left\{ \vec{r}_{i}\left( t\right) \right\} $ are the dynamical
variables and $E_{i}=\nu_{i}\left( \varepsilon/a\right) _{i}$ is the ``rest
energy'' of the particle $i.$ The SPH model action is then 
\begin{equation}
I_{SPH}=-\int dt\sum_{i}\left( \frac{E}{\gamma}\right) _{i}.
\label{Action_SPH}
\end{equation}

\subsection{Variational Procedure}

The Lagrangian, given by Eq.(\ref{L_SPH}), seems to show that the system is
as if it is just a sum of free particles of \ ``rest \ mass'' $E_{i}$.
However, there exists a basic difference. Namely, $E_{i}^{\text{ }}$'s are
not constant with respect to the variation of positions. This is because the
energy contained in each particle is determined by the configuration of
other particles through thermodynamical relations. More explicitly,
variations in the particles positions $\left\{ \vec{r}_{i},i=1,...,n\right\} 
$ cause change of the volume occupied by each particle, which in turn
modifies its energy (rest energy).

For our variational procedure, we consider changes of quantities associated
only kinematically with virtual variations of the configuration. That is,
except for the energy and the volume, all extensive quantities such as the
entropy and the particle number should be kept constant while variations of $%
\left\{ \vec{r}_{i}\right\} $ are taken. This implies that for variations of 
$\vec{r}_{i}$, we keep $\nu_{i}$ constant if $A$ is not the energy or the
volume. Now, if we take $A$ as the volume, then its changes associated to
variations of $\vec{r}_{i}$ are described by Eq.(\ref{omega_r}) where $\nu
_{i}$ should be understood as the initial volume of the particle $i$.
Therefore, even for the case $A$ is the volume, $\nu_{i}$ should be kept
constant for variations of $\vec{r}_{i}$. In short, except for the case of
the energy, the parameters $\nu_{i}$ should be kept constant while
variations of $\left\{ \vec{r}_{i}\right\} $ are taken. If we take $A$ as
the energy, then the change of $\nu_{i}$ should be calculated as a function
of the variation in $\left\{ \vec{r}_{i}\right\} $. This introduces an
additional complication without any practical merit. Therefore, for the sake
of simplicity, we restrict ourselves to the case where $A$ is not the energy
but any other extensive quantity.

We can write the change of energy associated with a virtual change of volume 
$\delta V$ as 
\begin{align}
\delta E & =-P\delta V+\delta W  \label{dE=-PdV} \\
& =-P_{eff}\delta V,
\end{align}
where $\delta W$ is an additional work to change irreversibly the volume and 
$P_{eff}$ is the effective pressure. If the change of volume is performed in
a quasi-static way and there exists no dissipative force, then this
effective pressure coincides with the usual pressure $P$. However, if there
exists some irreversible process associated with the volume change, then $%
P_{eff}\neq P$ and we may write 
\begin{equation}
P_{eff}=P+Q,  \label{Peff}
\end{equation}
where $Q\delta V$ is the energy change due to some irreversible process. For
a quasi-static adiabatic process, $Q=0.$ The introduction of $Q$ is
important when we deal with shock phenomena (see Sec.III).

The variation of volume $\delta V$ for constant $\nu_{i}$'s is calculated
from Eq.(\ref{V}) as 
\begin{align}
\delta V_{i} & =V_{i}\left( \frac{\delta\gamma_{i}}{\gamma_{i}}-\frac{\delta
a_{i}^{\ast}}{a_{i}^{\ast}}\right)  \nonumber \\
& =-\frac{\nu_{i}}{a_{i}}\left( -\gamma_{i}^{2}\vec{v}_{i}\cdot\delta\vec {v}%
_{i}+\frac{1}{a_{i}^{\ast}}\sum_{j}\nu_{j}\left( \delta\vec{r}_{i}-\delta%
\vec{r}_{j}\right) \cdot\nabla W_{ij}\right) ,
\end{align}
where $W_{ij}\equiv W\left( \vec{r}_{i}-\vec{r}_{j};h\right) $. Using these
relations, the variation of the action (\ref{Action_SPH}) with respect to $%
\left\{ \vec{r}_{i},i=1,...,n\right\} $ becomes 
\begin{align*}
\delta I_{SPH} & =-\int dt\sum_{i}\frac{1}{\gamma_{i}}\;\left( -\left(
P+Q\right) _{i}\delta V_{i}-E_{i}\frac{\delta\gamma_{i}}{\gamma_{i}}\right)
\\
& =-\int dt\sum_{i}d\vec{r}_{i}\cdot\left\{ \frac{d}{dt}\left[ \nu
_{i}\left( \frac{\varepsilon+P+Q}{a}\right) _{i}\gamma_{i}\vec{v}_{i}\right]
\right. \\
& \left. +\nu_{i}\sum_{j}\nu_{j}\left[ \frac{1}{\gamma_{i}^{2}}\left( \frac{%
P+Q}{a^{2}}\right) _{i}+\frac{1}{\gamma_{j}^{2}}\left( \frac {P+Q}{a^{2}}%
\right) _{j}\right] \nabla_{i}W_{ij}\right\} .
\end{align*}
Here, we used the symmetry of $W$ and the property 
\begin{equation}
\nabla_{i}W_{ij}=-\nabla_{j}W_{ij}.
\end{equation}
The requirement $\delta I_{SPH}=0$ for $\forall\delta\vec{r}_{a}$ leads to 
\begin{equation}
\frac{d}{dt}\left[ \nu_{i}\left( \frac{P+Q+\varepsilon}{a}\right)
_{i}\gamma_{i}\vec{v}_{i}\right] =-\sum_{j}\left[ \frac{\nu_{i}\nu_{j}}{%
\gamma_{i}^{2}}\left( \frac{P+Q}{a^{2}}\right) _{i}+\frac{\nu_{i}\nu_{j}}{%
\gamma_{j}^{2}}\left( \frac{P+Q}{a^{2}}\right) _{j}\right] \nabla _{i}W_{ij}.
\label{Mon_2}
\end{equation}
The corresponding hydrodynamic equation in the continuum limit\cite
{Variation} is 
\begin{equation}
\partial_{\mu}T^{\mu\nu}=\partial_{\mu}\Sigma^{\mu\nu},  \label{rel-hyd}
\end{equation}
where 
\begin{equation}
\Sigma^{\mu\nu}=Q\left[ u^{\mu}u^{\nu}-g^{\mu\nu}\right]  \label{q_visc}
\end{equation}
is the stress tensor for the bulk viscosity $Q$. This equation is the same
as those discussed in Refs.(\cite{Zurek,Siegler,Thacker}).

Equation (\ref{Mon_2}) should be complemented by an equation which expresses
the conservation of energy (and other thermodynamical quantities, if any
diffusion process is present). The energy conservation is written as 
\begin{equation}
\frac{dE_{i}}{dt}=-\left( P+Q\right) _{i}\frac{dV_{i}}{dt}.  \label{Cons_E}
\end{equation}
On the other hand, from the second law of thermodynamics, we should have 
\begin{equation}
\frac{dE_{i}}{dt}=-P\frac{dV_{i}}{dt}+T_{i}\frac{dS_{i}}{dt}+\mu_{i}\frac{%
dN_{i}}{dt}
\end{equation}
in equilibrium, where $T,S,\mu,N$ are the temperature, entropy, chemical
potential and particle number, respectively. Thus, we get 
\begin{equation}
T_{i}\frac{dS_{i}}{dt}+\mu_{i}\frac{dN_{i}}{dt}=-Q_{i}\frac{dV_{i}}{dt}.
\end{equation}
In the absence of particle diffusion, the chemical equilibrium requires 
\begin{equation}
\mu\frac{dN}{dt}=0
\end{equation}
and, in this case, 
\begin{equation}
T_{i}\frac{dS_{i}}{dt}=-Q_{i}\frac{dV_{i}}{dt}.  \label{DS}
\end{equation}

\subsection{Entropy Representation of SPH Equations}

As we have mentioned in the Introduction, in applications to
ultrarelativisitc nuclear collisions the baryon number is not a suitable
quantity to represent the hydrodynamic flow, since most of the energy
content is in the form of non-baryonic matter. This is particularly so in
the central rapidity region. We may consider the energy content itself as
the SPH base. However, as mentioned before, this choice introduces an
additional constraint between the coordinates $\left\{ \vec{r}_{i}\right\} $
and the extensive parameters $\left\{ \nu_{i}\right\} $ of SPH particles,
due to the energy conservation, and not desirable from the practical point
of view. Therefore, we propose to take the entropy as the suitable extensive
quantity for the SPH representation.

Let $s^{\ast}\left( t,\vec{r}\right) $ be the entropy density in the
space-fixed (calculational) frame. From Eq.(\ref{a*}), we have 
\begin{align*}
s_{i}^{\ast} & =s^{\ast}\left( t,\vec{r}_{i}\left( t\right) \right) \\
& =\sum_{j}\nu_{j}W\left( \vec{r}_{i}-\vec{r}_{j}\right) ,
\end{align*}
and the equations of motion for $\left\{ \vec{r}_{i}\right\} $ are given by
Eq.(\ref{Mon_2}), which we write in the form, 
\begin{align}
\frac{d\vec{r}_{i}}{dt} & =\vec{v}_{i,}  \label{a} \\
\frac{d\vec{\pi}_{i}}{dt} & =-\sum_{j}\left[ \frac{\nu_{i}\nu_{j}}{%
s_{i}^{\ast2}}\left( P+Q\right) _{i}+\frac{\nu_{i}\nu_{j}}{s_{j}^{\ast2}}%
\left( P+Q\right) _{j}\right] \nabla_{i}W_{ij},  \label{b}
\end{align}
where 
\begin{equation}
\vec{\pi}_{i}=\nu_{i}\left( \frac{P+Q+\varepsilon}{s}\right) _{i}\gamma _{i}%
\vec{v}_{i}  \label{c}
\end{equation}
is the momentum density associated with the $i-th$ particle.

For the variational procedure, $\nu _{i}$'s are kept constant but it does
not mean that they are constant in time. When there exists some
non-adiabatic process, then $Q\neq 0$ and we have to use the energy
conservation equation, Eq.(\ref{DS}) to determine the change of these
parameters $\left\{ \nu _{i}\right\} $. We have 
\begin{equation}
\frac{1}{\nu _{i}}\frac{d\nu _{i}}{dt}=-\frac{Q_{i}}{Ts_{i}^{\ast
}}\theta _{i},  \label{dnudt}
\end{equation}
where 
\begin{align}
\theta _{i}& =\frac{1}{V_{i}}\frac{dV_{i}}{dt}  \nonumber \\
& =\partial _{\mu }u^{\mu }.
\end{align}
Eqs. (\ref{a},\ref{b},\ref{dnudt}) constitute a system of first-order
differential equations for $\vec{r}_{i},\vec{\pi}_{i}$ and $\nu _{i}$, where
the velocity $\vec{v}_{i}$ should be determined algebraically from Eq.(\ref
{c}). Further, we have to specify the equation of state, for example, 
\[
E=E\left( N,S\right) ,
\]
and the dissipation pressure, 
\[
Q=Q(N,S,\theta ).
\]
We will discuss these points later in practical examples.

\subsection{SPH Equation for Generalized Coordinate System}

The variational procedure can readily be extended to coordinate systems with
non-Cartesian metric. The use of the generalized coordinate system is
particularly important when we consider realistic initial conditions for
simulations of RHIC processes. As we know, in a relativistic heavy-ion
collisional process, the initial state is a cold, quantum nuclear matters
flying separately. Just after the collision, the hadronic matter stays at a
highly excited state and the materialization occurs only after $\symbol{126}%
1fm/c$ in the proper time. Therefore, the local thermodynamical state would
emerge for some local proper time and not for the global space-fixed time $t$%
. Thus, it is important to choose a convenient coordinate system for the
description of relativistic heavy-ion collisions. For example, it is often
used the hyperbolic time and longitudinal coordinates as is described later.

Let us consider a more general coordinate system, 
\begin{equation}
ds^{2}=g_{\mu\nu}dx^{\mu}dx^{\nu}.
\end{equation}
However, in order to unambiguously define the conserved quantities, we
consider only the case when the time-like coordinate is orthogonal to the
space-like coordinates, 
\begin{equation}
g_{\mu0}=0.
\end{equation}
The action principle for the relativistic fluid motion can be written as\cite
{Variation} 
\begin{equation}
\delta I=-\delta\int d^{4}x\sqrt{-g}\varepsilon=0\,,
\end{equation}
together with the constraint for the conserved entropy current, 
\begin{equation}
\left( su^{\mu}\right) _{;\mu}=\frac{1}{\sqrt{-g}}\partial_{\mu}\left( \sqrt{%
-g}su^{\mu}\right) =0\,
\end{equation}
or 
\begin{equation}
\frac{1}{\sqrt{-g}}\partial_{\tau}\left( \sqrt{-g}s\gamma\right) +\frac {1}{%
\sqrt{-g}}\sum_{i}\partial_{i}\left( \sqrt{-g}s\gamma v^{i}\right) =0\,,
\end{equation}
where 
\begin{equation}
v^{i}=\frac{u^{i}}{u^{0}}\;
\end{equation}
and we use the notation, 
\[
\tau=x^{0},\,\;\gamma=u^{0}. 
\]
The generalized\ gamma factor $\gamma$ is related to the velocity $\vec{v}%
_{a}$ through $u_{\mu}u^{\mu}=1,$ so that 
\begin{equation}
\gamma=\frac{1}{\sqrt{g_{00}-\vec{v}^{T}{\bf g}\vec{v}}}\,,  \label{4}
\end{equation}
where $-{\bf g}$ is the $3\times3$ space part of the metric tensor. That is 
\begin{equation}
\left( g_{\mu\nu}\right) =\left( 
\begin{array}{cc}
g_{00} & 0 \\ 
0 & -{\bf g}
\end{array}
\right) \,.
\end{equation}

Let us now introduce the SPH representation. We may, for example, express
the entropy density by the ansatz 
\begin{equation}
\sqrt{-g}s\gamma=s^{\ast}\rightarrow s_{SPH}^{\ast}=\sum_{i}\nu_{i}W\left( 
\vec{r}-\vec{r}_{i}\left( \tau\right) \right) ,  \label{sph_1}
\end{equation}
or by 
\begin{equation}
s\gamma=s^{\ast}\rightarrow s_{SPH}^{\ast}=\sum_{i}\nu_{i}W\left( \vec {r}-%
\vec{r}_{i}\left( \tau\right) \right) ,  \label{sph_2}
\end{equation}
as well. These two possibilities, besides others, are simply different ways
to parametrize a variational ansatz in terms of a linear combination of
given functions $W\left( \vec{r}-\vec{r}_{i}\left( \tau\right) \right) $.
The most important property of an ansatz should be that $W$ satisfies the
normalization condition imposed by the basic conserved quantity. Since the
total entropy is expressed as 
\begin{equation}
S=\int d^{3}\vec{r}\sqrt{-g}s\gamma=\sum_{i}\nu_{i}\,,
\end{equation}
the normalization of $W$ should be taken to be 
\begin{equation}
\int d^{3}\vec{r}W\left( \vec{r}-\vec{r}\,^{\prime}\right) =1,
\label{norm_1}
\end{equation}
for the parametrization Eq.(\ref{sph_1}) and 
\begin{equation}
\int d^{3}\vec{r}\sqrt{-g}W\left( \vec{r}-\vec{r}\,^{\prime}\right) =1,
\label{norm_2}
\end{equation}
for the parametrization Eq.(\ref{sph_2}). In the usual SPH calculations, it
is not desirable to introduce in $W$ the space-time dependence through its
normalization condition. In this aspect, the most natural way to introduce
the SPH representation is Eq.(\ref{sph_1}). With this choice, the SPH action
is given by

\begin{align}
I_{SPH} & =-\int d\tau\int d^{3}\vec{x}\sum_{i}\nu_{i}\left( \frac{\sqrt {-g}%
\varepsilon}{\sqrt{-g}s\gamma}\right) _{i}W\left( \vec{r}-\vec{r}_{i}\left(
\tau\right) \right)  \nonumber \\
& =-\int d\tau\sum_{i}\nu_{i}\left( \frac{\varepsilon}{s\gamma}\right) _{i}.
\label{SPH_action}
\end{align}
The variational principle leads to the following equation of motion, 
\begin{align}
\frac{d}{d\tau}\vec{\pi}_{i} & =-\sum_{j}\nu_{i}\nu_{j}\left[ \frac{1}{\sqrt{%
-g_{i}}\gamma_{i}^{2}}\frac{P_{i}+Q_{i}}{s_{i}^{2}}+\frac{1}{\sqrt{-g_{j}}%
\gamma_{j}^{2}}\frac{P_{j}+Q_{j}}{s_{j}^{2}}\right] \nabla _{i}W_{ij} 
\nonumber \\
& +\frac{\nu_{i}}{\gamma_{i}}\frac{P_{i}+Q_{i}}{s_{i}}\left( \frac{1}{\sqrt{%
-g}}\nabla\sqrt{-g}\right) _{i}  \nonumber \\
& +\frac{\nu_{i}}{2}\gamma_{i}\left( \frac{P+Q+\varepsilon}{s}\right)
_{i}\left( \nabla g_{00}-\vec{v}_{i}^{T}\nabla{\bf g}\vec{v}_{i}\right) _{i},
\label{EQM}
\end{align}
where 
\begin{equation}
\vec{\pi}_{i}=\gamma_{i}\nu_{i}\left( \frac{P+Q+\varepsilon}{s}\right) _{i}%
{\bf g}\vec{v}_{i}  \label{q}
\end{equation}
and the operator $\nabla$ is just the simple derivative operator with
respect to the coordinate variable in use.

\subsection{Hyperbolic Coordinates}

For ultrarelativistic heavy-ion collisions, a useful set of variables are 
\begin{align}
\tau & =\sqrt{t^{2}-z^{2}},  \label{tau} \\
\eta & =\frac{1}{2}\tanh\frac{t+z}{t-z},  \label{eta} \\
\vec{r}_{T} & =\left( 
\begin{array}{c}
x \\ 
y
\end{array}
\right) .
\end{align}
As mentioned above, the initial conditions for RHIC processes are specified
in terms of proper time rather than of coordinate time $t$. The variable $%
\tau$ is not exactly the physical proper time of the matter, but for the
initial times it may approximate the proper time.

The metric tensor for this coordinate system is given by 
\begin{align*}
g_{00} & =1, \\
{\bf g} & =\left( 
\begin{array}{ccc}
1 & 0 & 0 \\ 
0 & 1 & 0 \\ 
0 & 0 & \tau^2 
\end{array}
\right) , \\
\sqrt{-g} & =\tau.
\end{align*}
Since the metric is space independent, we can use the parametrization, 
\[
\tau\gamma_{i}s_{i}=s_{i}^{\ast}=\sum_{j=1}^{n}\nu_{j}W\left( q_{ij}\right)
, 
\]
where 
\[
q_{ij}=\sqrt{\left( x_{i}-x_{j}\right) ^{2}+\left( y_{i}-y_{j}\right)
^{2}+\tau^{2}\left( \eta_{i}-\eta_{j}\right) ^{2}} 
\]
and $W$ is normalized as 
\[
4\pi\int_{0}^{\infty}q^{2}dq\;W\left( q\right) =1. 
\]
The SPH equation becomes 
\[
\frac{d}{d\tau}\vec{\pi}_{i}=-\frac{1}{\tau}\sum_{j}\nu_{i}\nu_{j}\left[ 
\frac{1}{\gamma_{i}^{2}}\frac{P_{i}+Q_{i}}{s_{i}^{2}}+\frac{1}{\gamma_{j}^{2}%
}\frac{P_{j}+Q_{j}}{s_{j}^{2}}\right] \nabla_{i}W_{ij}\,, 
\]
where the $\eta$ component of the momentum is related to the velocity $%
d\eta/d\tau$ as 
\[
\pi_{\eta}=\tau^{2}\nu\gamma\left( \frac{P+Q+\varepsilon}{s}\right) \frac{%
d\eta}{d\tau}, 
\]
whereas in the transverse directions, we have 
\[
\vec{\pi}_{T}=\nu\gamma\left( \frac{P+Q+\varepsilon}{s}\right) \frac {d\vec{r%
}_{T}}{d\tau}. 
\]
The Lorentz factor is given by 
\[
\gamma=\frac{1}{\sqrt{1-\vec{v}_{T}^{2}-\tau^{2}v_{\eta}^{2}}}\,. 
\]

\section{Examples}

We have formulated the relativistic SPH method, appropriate to the study of
\ RHIC processes. In order to check its validity and efficiency we apply, in
this section, our method to several known problems and compare the results
to the analytic or known numerical solutions.

\subsection{Adiabatic Flow of Massless Pion Gas}

For adiabatic flows of perfect gas, the entropy is conserved. Thus $\dot{\nu 
}_{i}=0$ and we have $Q=0$. In the following examples, we consider the cases
where $Q=0.$

\subsubsection{Landau Model}

Let us first investigate a well-known analytically soluble flow. The Landau
model is one of the few examples of relativistic fluid flow for which
analytical solution is available. Consider a one-dimensional, relativistic,
massless, baryon-free gas initially at rest. The equation of state is 
\[
P=\frac{1}{3}\varepsilon=Cs^{4/3}, 
\]
where 
\[
C=\left( \frac{15}{128\pi^{2}}\right) ^{1/3}. 
\]
Since we are dealing with a perfect gas, $Q=0$. To apply the SPH method, we
introduce the discrete one dimensional space variable $\left\{ \eta
_{i}\left( t\right) ,i=1,..,n\right\} $. The relation between the momentum
and velocity is then 
\begin{equation}
\pi=C\nu s^{\ast1/3}\gamma^{2/3}v.  \label{q(v)}
\end{equation}
In this case, $v$ can be solved analytically in terms of $\pi$.

In Fig.1-a and -b, we show the results of our SPH calculations, together
with the exact solution \cite{Landau}.

\bigskip

Fig. 1-a, Fig.1-b

In this example, we took only $100$ particles with equally spaced $\left\{
\eta_{i}\right\} $. As we see, in spite of a rather small number of
particles, the SPH solution is quite satisfactory in this example. In
particular, when we use the $\eta-\tau$ coordinates with an appropriate
distribution of $\nu_{i}$'s (Fig. 1-b), an excellent agreement with the
analytical solution can be obtained.

\subsubsection{3D Scaling Solution}

A simple analytical solution for a 3 dimensional relativistic pion gas is
available. It is just a generalization of the one-dimensional scaling
solution and given by 
\[
s=\frac{s_{0}}{\sqrt{\tau^{2}-x^{2}-y^{2}}}. 
\]
To see the efficiency of the SPH approach presented here, we reproduce this
solution numerically in the full $3D$ numerical code, without making use of
the spherical symmetry. In Fig.2, we show the result of such calculations.

\bigskip Fig.2

As we see, our numerical calculations reproduces the analytical solution
fairly well. One of the advantages of the SPH approach is that the coding
for $3D$ cases is almost the same as for the $1D$ case. The only problem is
a rapid increases of the number of particles for higher dimensions if we
want to keep a high accuracy. A direct coding would require a computational
time proportional to $n^{2}$, where $n$ is the number of particles. However,
in the absence of long range forces such as the gravitational or Coulomb
interactions, we may apply techniques such as the linked-list method to
reduce the computational time to the order of $n\log n$. In the example
shown above, we used $50\times50\times50$ particles, which took less than 2
minutes for one time step in a reasonable PC(Pentium-II). Here, we put a
rather large number of particles, since it was a somewhat stringent test due
to the divergent nature of the solution at the border.

\subsubsection{Transverse Expansion on Longitudinal Scaling Expansion}

As a further test, closer to a realistic situation than that of Fig. 2, we
calculated the transverse expansion of a cylindrically symmetric homogeneous
massless pion gas, undergoing a longitudinal scaling expansion, and
initially at rest in transverse directions. In Fig. 3, we show an example.
Such a problem has been discussed by several authors as a useful base to
understand the transverse expansion. Here, we compare our results (again a
full $3D$ calculation without assuming cylindrical symmetry) with (2+1)
numerical results, obtained by the use of the method of characteristics.\cite
{Pottag}

Fig. 3

In this example, we used also $50\times50\times50$ particles. The result is
quite satisfactory. If we may decrease the accuracy by $10\%$, we can reduce
the particle number almost by one order of magnitude.

\subsection{Non Adiabatic Case: Shock Waves and Artificial Viscosity}

As seen in the previous examples, our entropy-based relativistic SPH method
works quite well for the adiabatic dynamics ($Q=0$) of the massless pion
gas. However, for the application to realistic problems, it is fundamental
to see how this scheme works for non-adiabatic cases ($Q\neq0$), too. For
this purpose, we study some examples of one dimensional shock problems.

\subsubsection{Compression Shock (Normal Matter)}

Whenever there exists a shock wave, always exists the production of entropy
through the shock front. The shock front manifests as a discontinuity in
thermodynamical quantities in a hydrodynamic solution. Mathematically
speaking, the shock front should be treated as a boundary to connect two
distinct hydrodynamic solutions. To reproduce such a discontinuous behavior,
the full degrees of freedom of hydrodynamics are required. The smoothed
particle ansatz excludes such a possibility from the beginning. Since there
do not exist short-wavelength excitation modes in the SPH ansatz, the energy
and momentum conservation required by the hydrodynamics results in very
rapidly oscillating motion of each particle. Such a situation occurs, for
example, when a very high energy density gas is released into a low density
region. This kind of shock, for the case of a baryon gas, is discussed in
Ref.(\cite{Maruhn}) and also, in the SPH context, in Ref.(\cite{Siegler}).
Here, we applied our entropy-based SPH approach to the massless pion gas.

Fig.4 gives a typical behavior of SPH solution for such a situation, if
entropy production is not taken into account. As discussed above, there
appear rapid oscillations in thermodynamical quantities just behind the
shock front. Actually, this oscillation appears always in numerical
approaches if entropy production is not included.

In order to avoid these oscillations, von Neuman and Richtmeyer\cite{Neuman}
introduced the concept of pseudoviscosity. The idea is to set the
dissipative pressure where the shock wave discontinuity is present. To do
this, Neuman and Richtmeyer proposed the ansatz 
\[
Q=\left\{ 
\begin{array}{ccc}
\left( \alpha\Delta x\right) ^{2}\rho\;\left( \dot{\rho}/\rho\right) ^{2}, & 
\;\;\; & \dot{\rho}>0 \\ 
0,\;\;\;\;\;\;\;\;\;\;\;\; & \;\;\; & \;\dot{\rho}<0
\end{array}
\right. 
\]
for nonrelativistic one-dimensional hydrodynamics. Here, $\rho$ is the mass
density, $\Delta x$ is the space grid size and $\alpha$ is a constant of the
order of unity. In order to generalize the above pseudoviscosity for
relativistic SPH case, we replace the quantity $\dot{\rho}/\rho\,$ by $%
-\theta=-\partial_{\mu}u^{\mu}$ and $\Delta x$ by $h$, where $h$ is as
before the width of the smoothing kernel $W$. More precisely, we take the
following form which is a slightly modified expression suggested by Ref.( 
\cite{Siegler}), 
\begin{equation}
Q=\left\{ 
\begin{array}{ccc}
P\left[ -\alpha h\theta+\beta\left( h\theta\right) ^{2}\right] \,, & \;\;\;
& \theta<0\,, \\ 
0\,,\;\;\;\;\;\;\;\;\;\;\;\; & \;\;\; & \theta\geq0\,.
\end{array}
\right.  \label{Q}
\end{equation}

As mentioned before, $Q$ is equivalent to the bulk viscosity and therefore
there is no heat flow associated with it. What this artificial viscosity
does is to convert the collective flow energy into the microscopic thermal
energy. As a consequence, the total energy, that is, the sum of the
collective flow energy and the internal thermal energy is still conserved.

Fig.5 is the solution of the same problem of Fig.4, but with the entropy
production taken into account. In this calculation, the parameters have been
chosen as 
\[
\alpha =2,\;\beta =4 
\]
and $h=0.5fm$ for $1000$\thinspace SPH particles. As we see, the rapid
oscillations have been smoothed out (and in turn, the numerical calculation
became much more efficient).

It is known that the overall energy- and momentum-flux conservation relates
the ratio $s_{2}/s_{1}$ of entropy densities after and before the shock to
the velocity $v_{s}$ of the shock front as (Hugoniot-Rankine relation) 
\begin{equation}
\frac{s_{2}}{s_{1}}=\frac{2}{3^{3/4}}v_{s}\frac{\left( 9v_{s}^{2}-1\right)
^{1/4}}{\left( 1-v_{s}^{2}\right) ^{5/4}}\,.  \label{H-R}
\end{equation}
In Fig. 6, we show the velocity of the shock front obtained in our SPH
calculations as function of the entropy ratio(dots). Each point corresponds
to the different initial condition. They are compared with the
Hugoniot-Rankine relation Eq.(\ref{H-R}) (curve). The accordance shows that
our SPH calculation reproduces faithfully the conservation of kinetic energy
and momentum of the flow through the shock front.

\subsubsection{Rarefaction Shock}

When the fluid presents a first order phase transition, a discontinuity
appears in an expansion regime. This kind of shock wave has been discussed
in connection to the QGP-hadron phase transition\cite{Blaizot,Baym,Ruusk}.
In the present example, we use a simple MIT bag-model equation of state, 
\[
P=\left\{ 
\begin{array}{ccc}
\frac{\pi^{2}}{30}T^{4}, & \;\;\; & T\leq T_{c}\,, \\ 
\frac{37}{3}\frac{\pi^{2}}{30}T^{4}-B, & \;\;\; & T\geq T_{c}\,,
\end{array}
\right. 
\]
being $B$ the bag constant and $T_{c}$ the critical temperature.

Because decompression occurs with a constant pressure, we should have
negative $Q$ values for $\theta>0$. We can use a similar expression for $Q$
as before, 
\[
Q=\left\{ 
\begin{array}{ccc}
0, & \;\;\; & \theta<0\,, \\ 
P_{c}\Theta\left[ \left( \varepsilon-\varepsilon_{h}\right) \left(
\varepsilon_{QGP}-\varepsilon\right) \right] \left[ -\alpha h\theta
-\beta\left( h\theta\right) ^{2}\right] , & \;\;\; & \theta\geq0\,,
\end{array}
\right. 
\]
where $\Theta$ represents the Heaviside step function to let the viscosity
be effective only in the transition region. In Fig. 6, we show a result of
our calculation and compare it to the analytic solution. The real sharp
shock front is a little bit smoothed, but this is due to the small number of
SPH particles ($N=1000$ in this example) and hence a large $h$. Using the
pseudoviscosity, the shock width turns out to be a few times $h$.

\section{Discussion and Perspectives}

In the usual hydrodynamic computations using space grids, the symmetry of
the problem is often a crucial factor to perform a calculation of reasonable
size. The SPH method cures this aspect and furnishes a robust algorithm
particularly appropriate to the description of processes where rapid
expansions of the fluid should be treated. In this paper, we formulated an
entropy-based SPH description of the relativistic hydrodynamics. We have
shown that this approach is very promising for the study of
ultrarelativistic nucleus-nucleus collision processes. The equations of
motion are derived by a variational procedure from the SPH model action with
respect to the Lagrangian comoving coordinates. This guarantees that the
method furnishes the maximal efficiency for a given number of degrees of
freedom, keeping strictly the energy and momentum conservation. For this
reason, solutions can be obtained with a very reasonable precision, with a
relatively small number of SPH particles. This is the basic advantage of the
present method, when we analyze the event-by-event dynamics of the
relativistic heavy-ion collisions.

On the other hand, the precision of this method increases rather slowly with
the number of SPH particles. Therefore, a relatively large number of
particles is required if one wants a very precise numerical solution.
However, for the application to the RHIC physics, we may need rather crude
precision especially if we consider the dubious validity of the rigorous
hydrodynamics. For a calculation with typically $10\%$ errors, the SPH
algorithm presented here furnishes a very efficient tool to study the flow
phenomena in the RHIC physics.

A fundamental difficulty of the relativistic hydrodynamics for viscous fluid 
\cite{Israel,Hiscock} is that the dissipation term causes an intrinsic
instability to the system described by Eq.(\ref{rel-hyd}). This instability
basically comes from the fact that the dissipation term contains $\theta$
(see Eqs.(\ref{q_visc},\ref{dnudt})), so that it necessarily introduces the
third time-derivative into the equation. This means that we have to specify,
at least, a part of the acceleration as the initial condition. Even we
specify the initial acceleration, the requirement of the internal
self-consistency among the equations above leads to intrinsically unstable
solutions. Israel proposed\cite{Israel,Hiscock} to cure these difficulties
by introducing higher-order thermodynamics with respect to deviations from
the equilibrium. In the examples presented in the present paper, we did not
address this question and simply estimated the quantity $\theta$ from the
quantities one time step before. In practice, this will cause no numerical
instability and the behavior of the solution is quite satisfactory.

For the future application of the present program, we need to specify more
realistic initial conditions and also to relate the final state to the
physical observable quantities, such as particle spectra. The first point is
now being in progress by introducing the initial energy and momentum
distributions of the matter using the NEXUS algorithm. As for the second
point, that is, the problem of particle production, the possibility of
incorporation of the continuous emission mechanism \cite{Grassi} is being
studied.

\bigskip

This work was supported in part by PRONEX (contract no. 41.96.0886.00),
FAPESP (contract nos. 98/02249-4 and 98/00317-2), FAPERJ (contract
no.E-26/150.942/99) and CNPq-Brasil.

\newpage

\begin{center}
{\bf FIGURE\ CAPTIONS}
\end{center}

\begin{description}
\item[Fig. 1:]  a) Entropy profiles of Landau Model for different times.
The numerical solutions correspond to the equal $\nu$ for all the particles
and the Cartesian coordinate system was used. b) The same but the hyperbolic
coordinate system was used (see text).

\item[Fig.2:]  Three dimensional scaling solution for the massless pion gas.
Cartesian coordinates are used for the transverse direction and the
hyperbolic coordinates ($\eta-\tau$) for the longitudial
direction. Despite the spherical symmetry, the SPH calculation has 
been carried out in the full 3D code.

\item[Fig. 3:]  Transverse temperature profiles of a cilyndrically symmetric
flow with longitudinally scaling expansion. The SPH results (circles) are
compared with the numerical solution of the space-fixed grid method. The SPH
calculation has been done in the full 3D code.

\item[Fig. 4:]  Shock wave in one dimensional pion gas. No viscosity is used.

\item[Fig. 5:]  After the introduction of the $Q$ term in the SPH
calculation.

\item[Fig. 6:]  Test of Hugoniot-Rankine relation. Black circles are those
of the SPH calculations with different initial conditions, and the solid
line is Eq.(\ref{H-R}).

\item[Fig. 7:]  Comparison of the SPH results with the analytical solution
for the rarefaction shock. Here we took $B=400MeV/fm^{3}$.
\end{description}

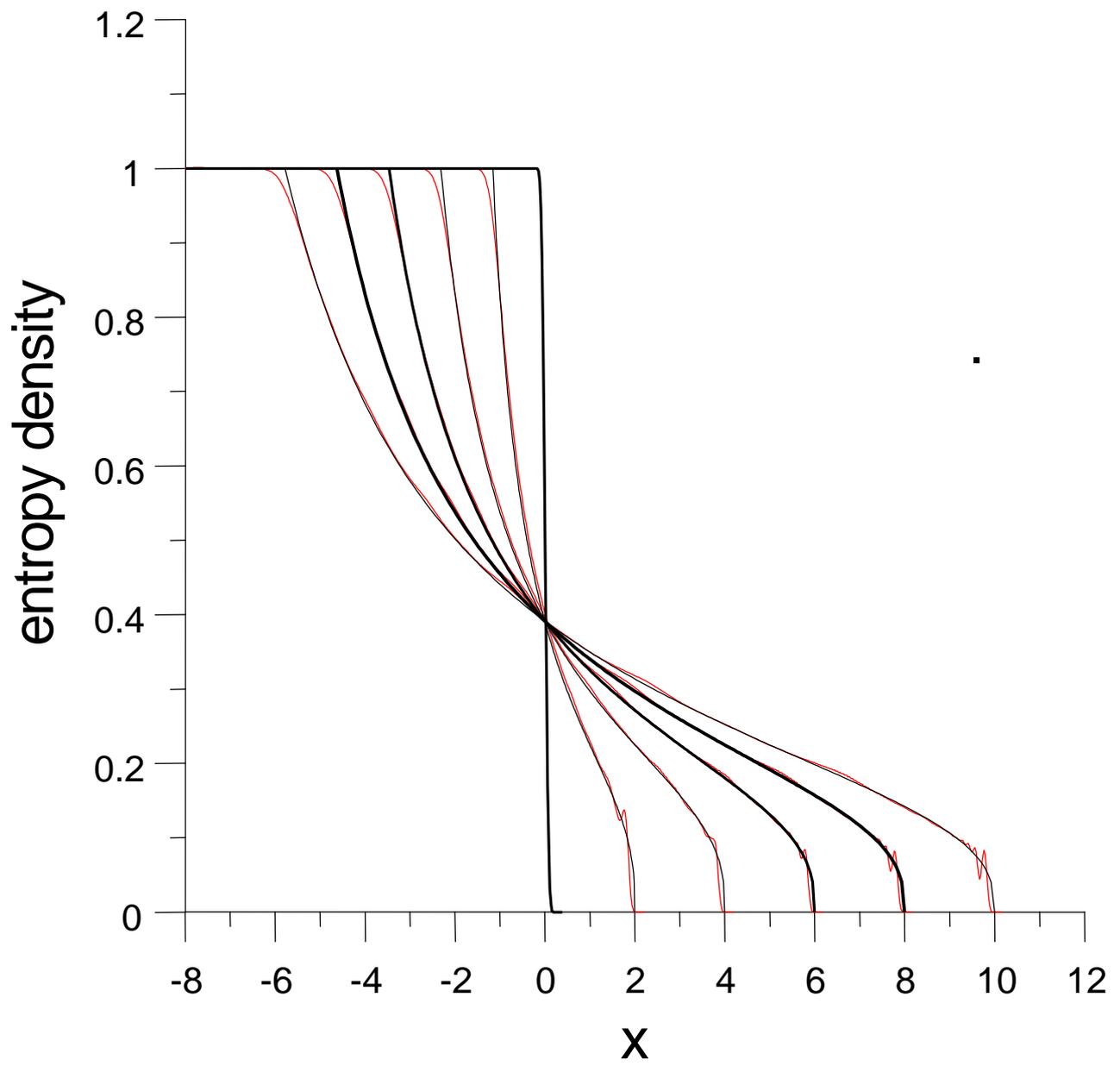
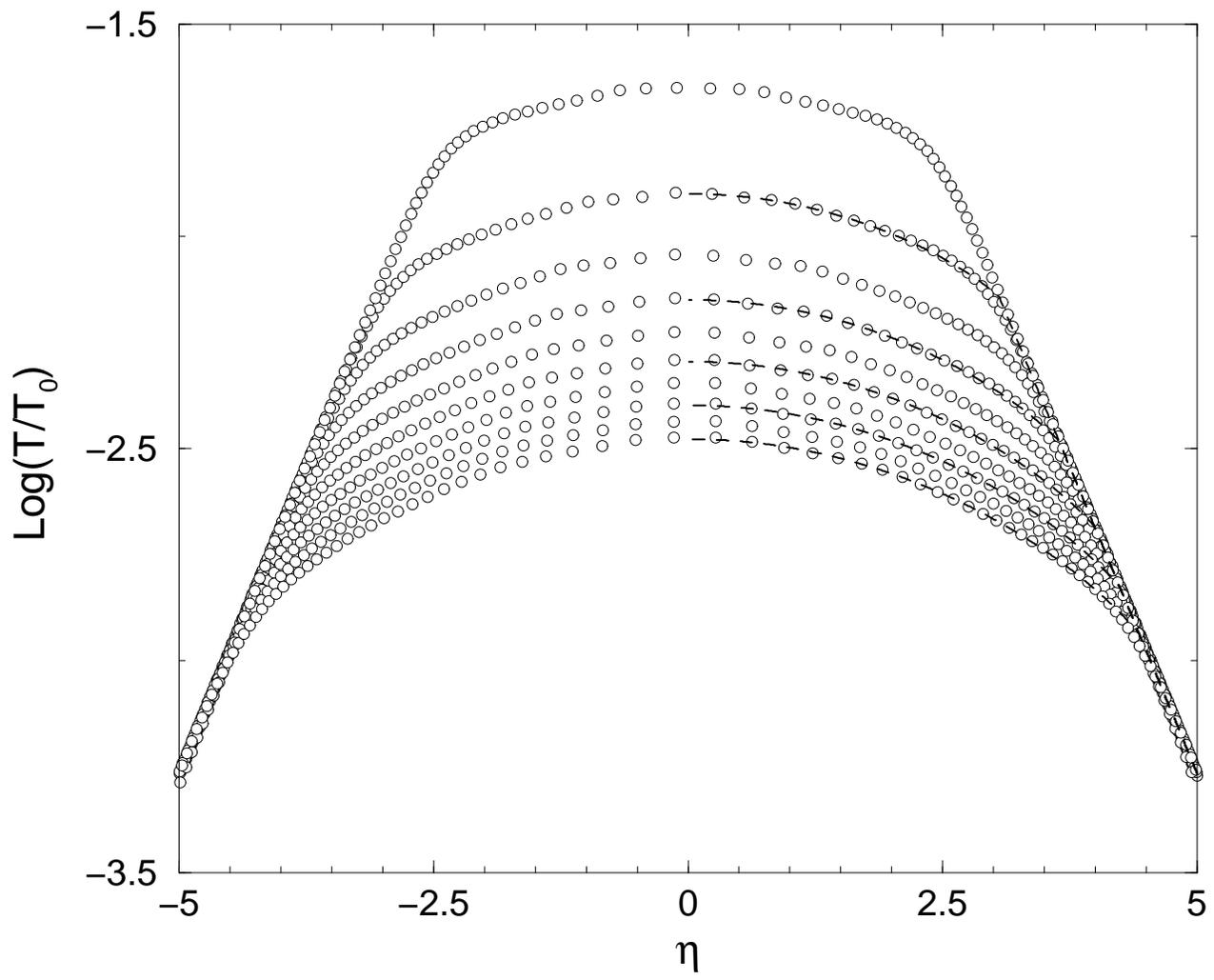
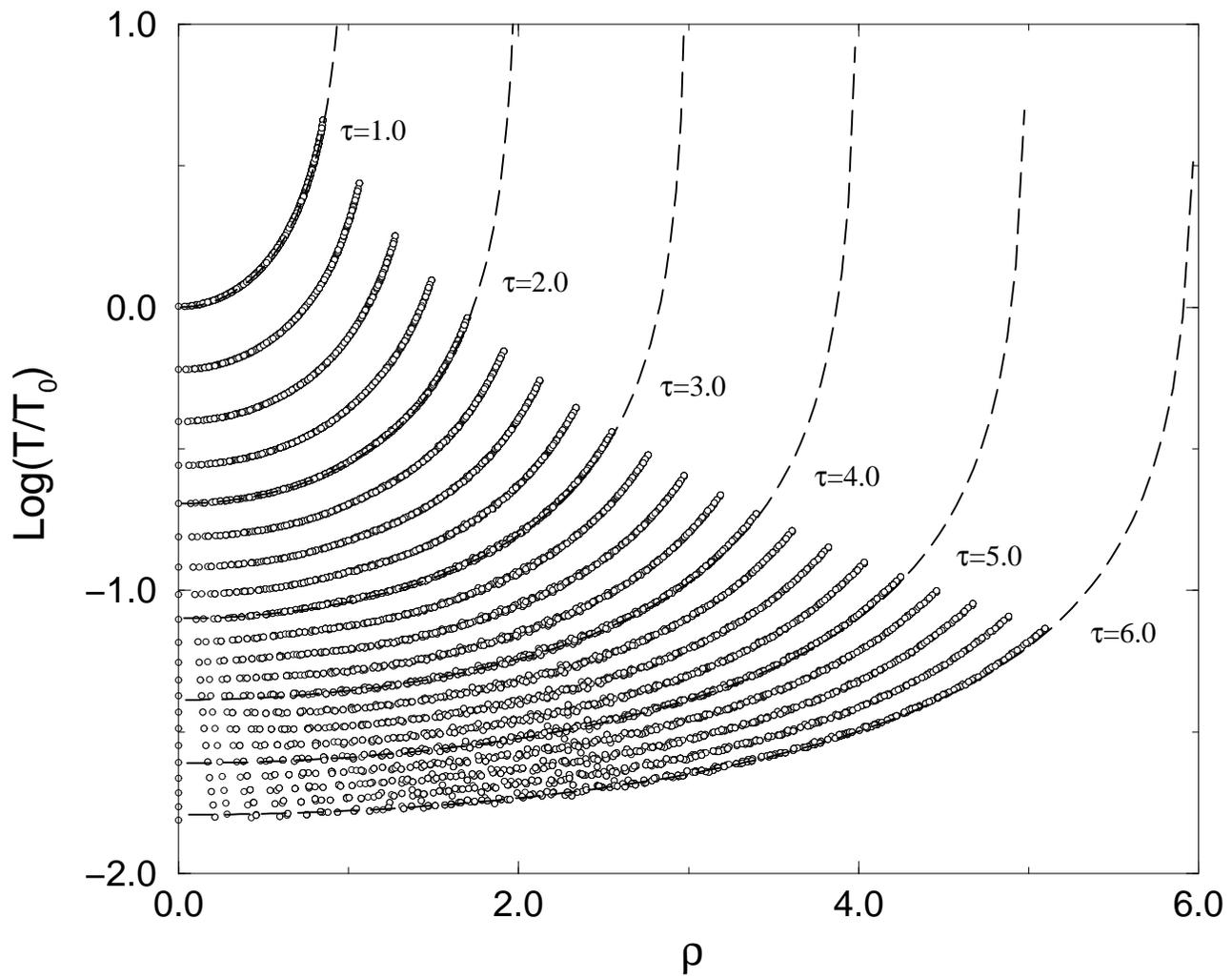
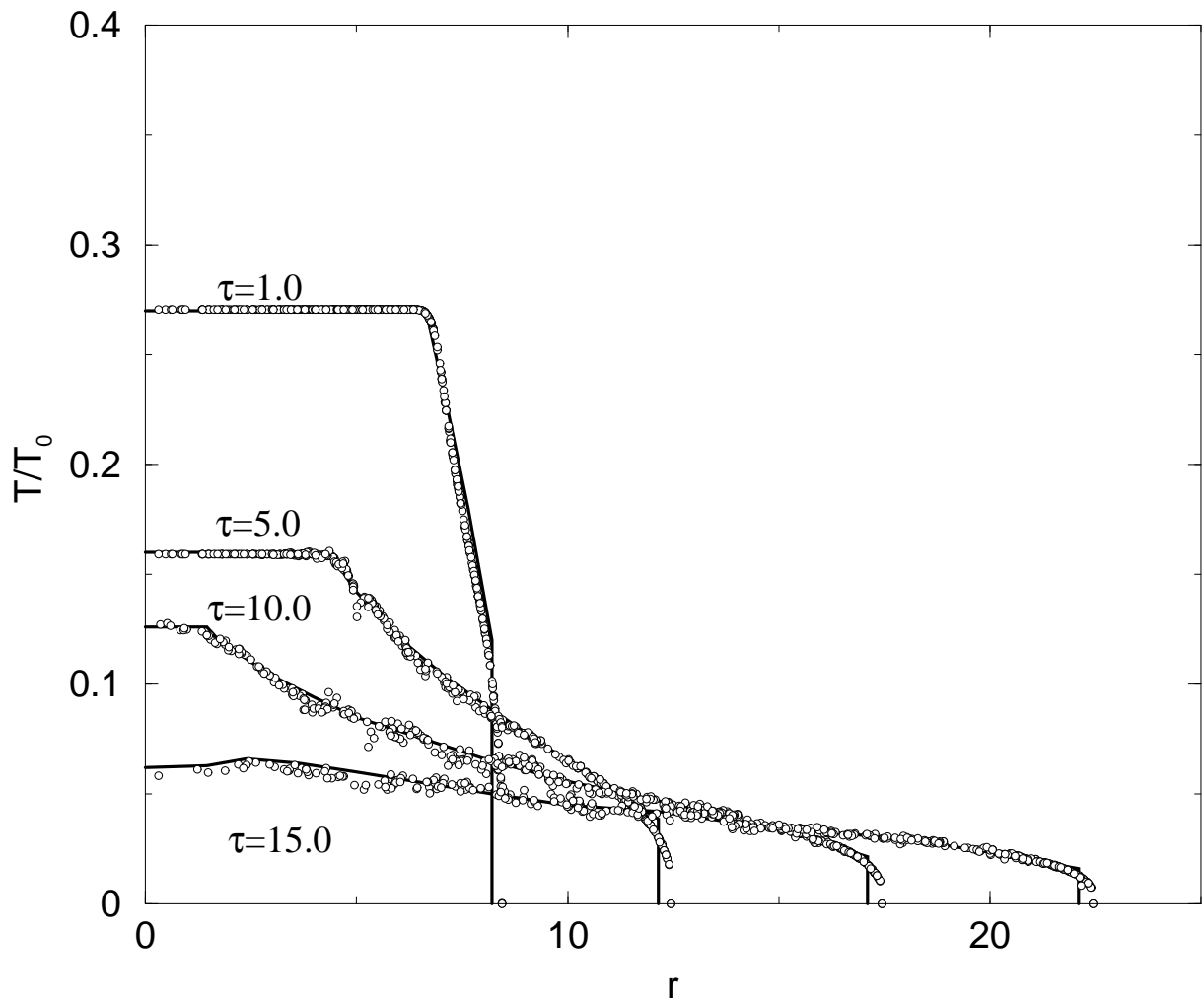
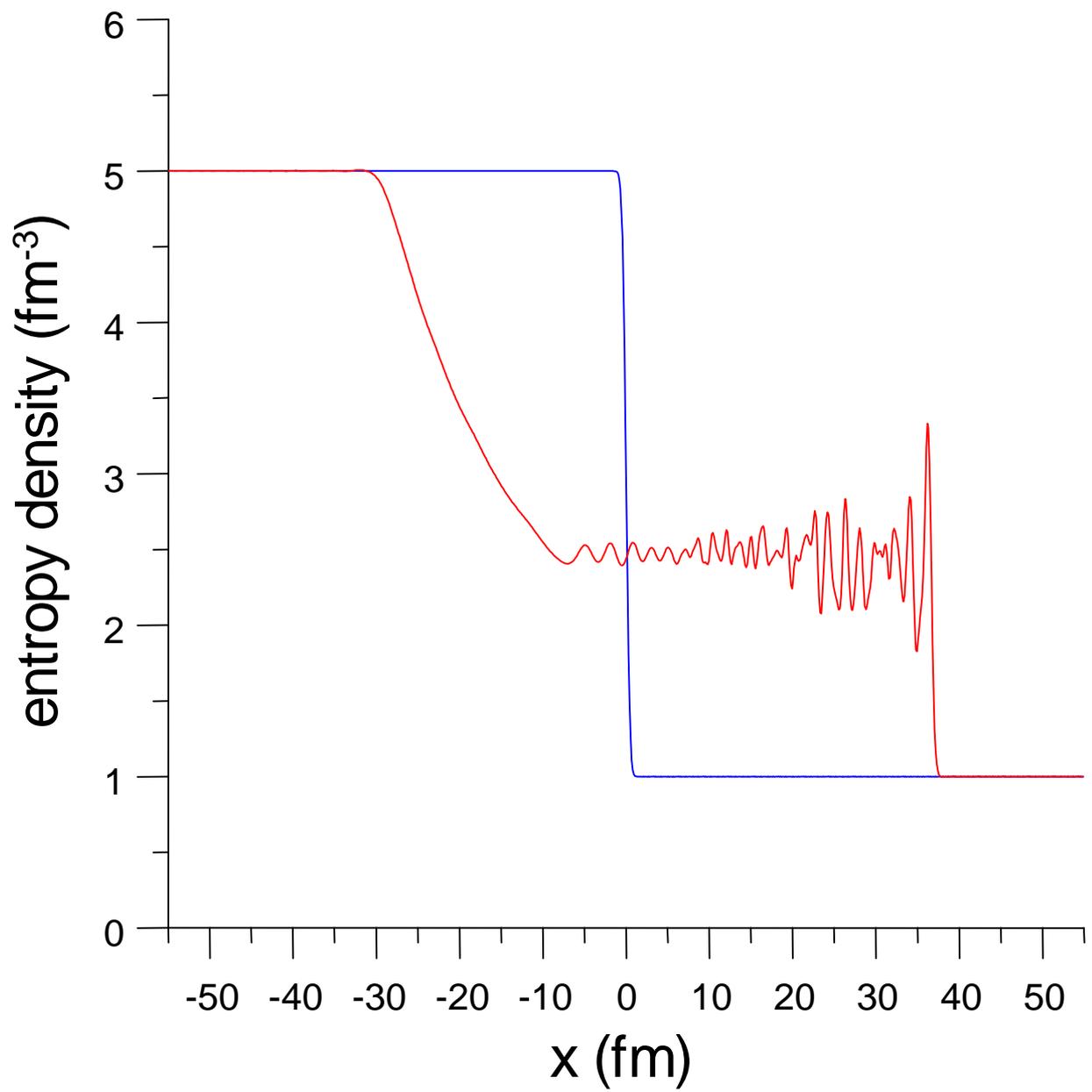
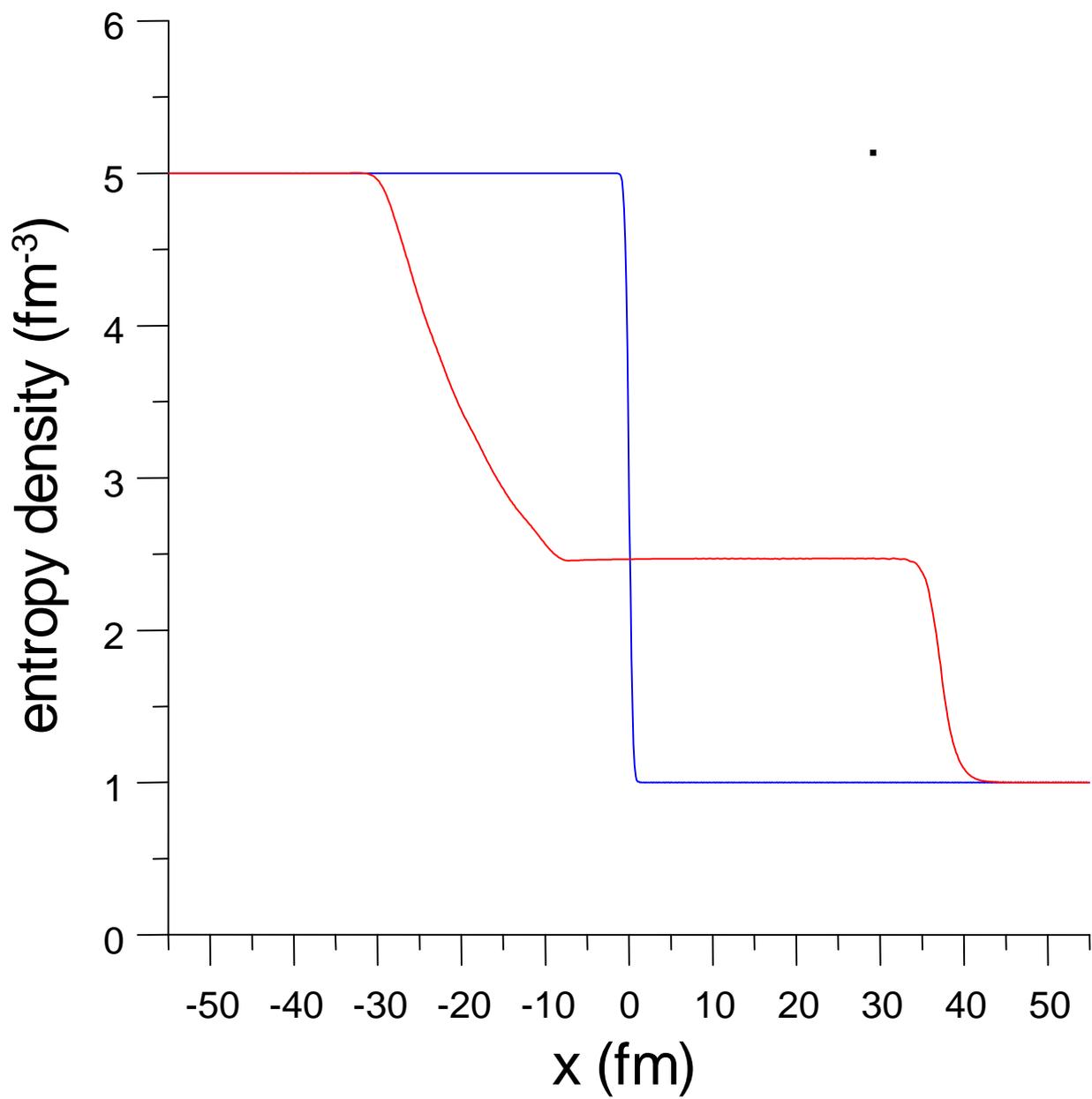
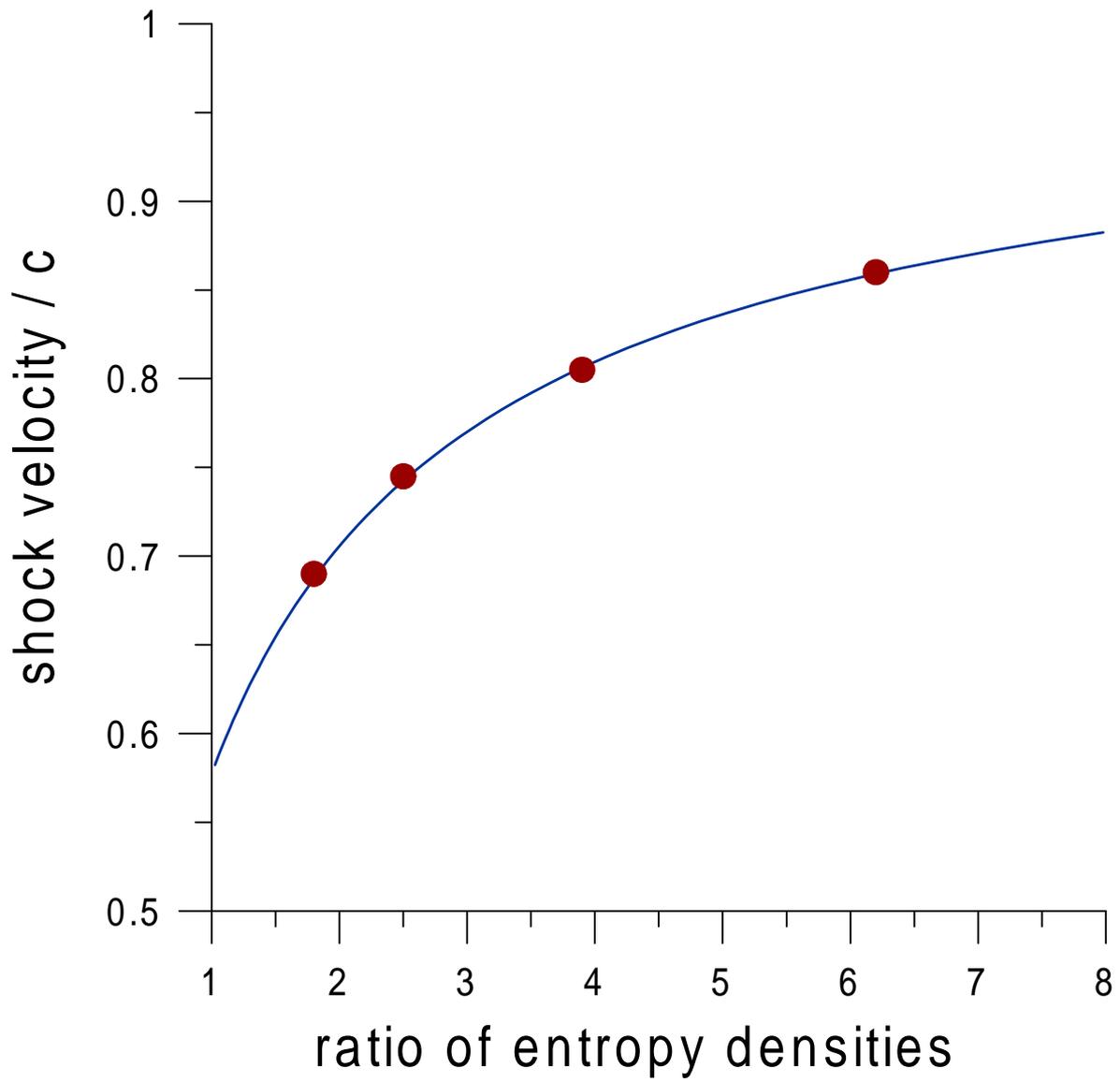
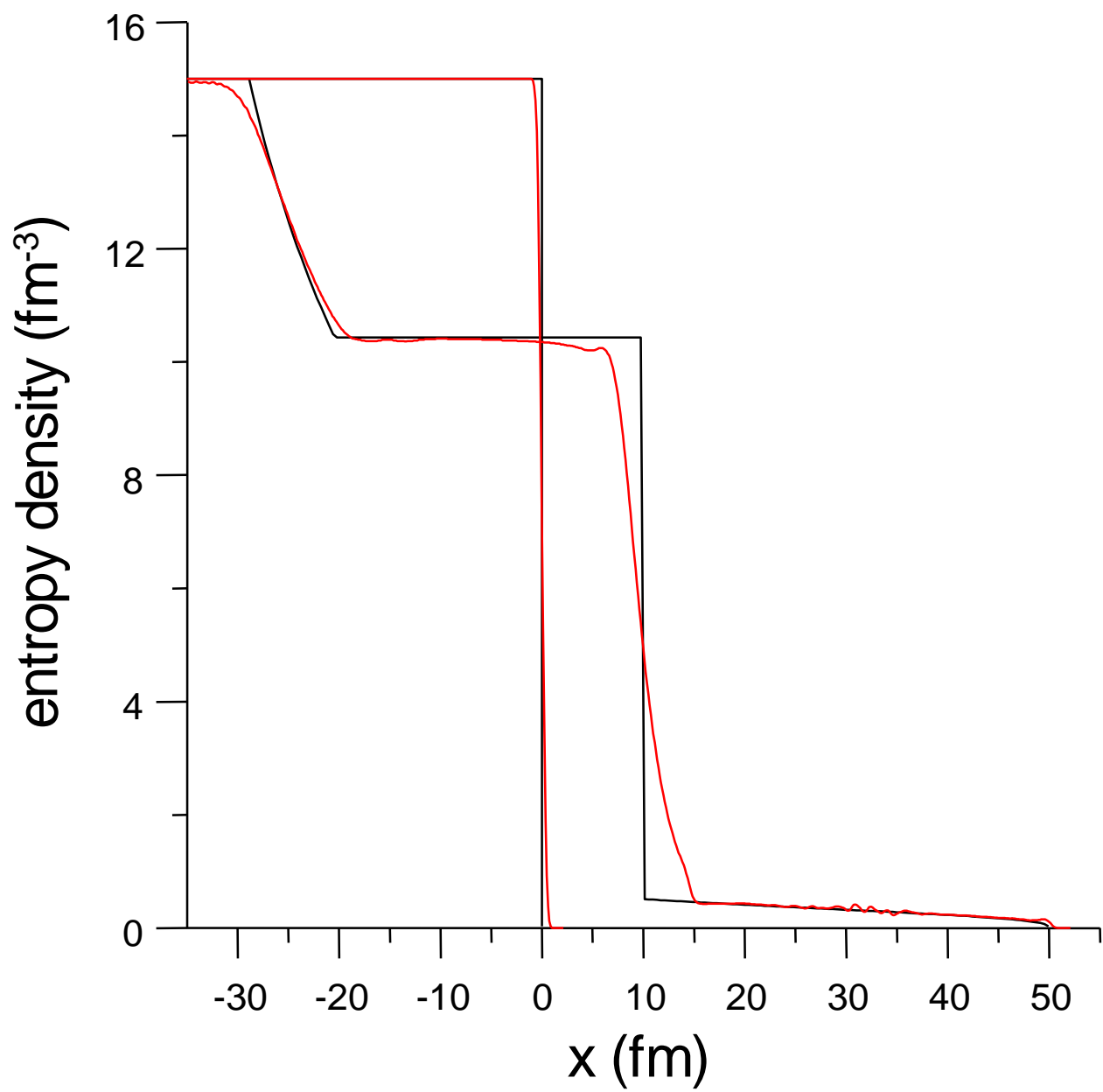
\end{document}